\documentclass[letter,prd,aps,twocolumn,floatfix,superscriptaddress]{revtex4}

\usepackage{amssymb}
\usepackage{graphicx,float,amsmath}
\usepackage{amssymb,amsmath,graphicx}
\usepackage{hyperref}
\usepackage[usenames,dvipsnames]{color}
\usepackage{array}

\begin{document}

\title{Towards the nonlinear regime in extensions to GR: assessing possible options}
\author{Gwyneth Allwright}
\affiliation{Perimeter Institute for Theoretical Physics, Waterloo, ON, N2L2Y5, 
Canada}
\author{Luis Lehner}
\affiliation{Perimeter Institute for Theoretical Physics, Waterloo, ON, N2L2Y5, 
Canada}

\begin{abstract}
Testing General Relativity and exploring possible departures has received further
input with the possibility to do so through gravitational waves emitted in 
strongly gravitating/highly dynamical scenarios and also through the availability of
exquisitely sensitive cosmological observations. However, most extensions suffer
from severe pathologies at the mathematical level which have stymied a thorough exploration
of putative theories. With the aid of a model problem which captures typical pathologies, we explore
suggested methods to control them. We find that the approach that modifies the equations to
control higher-order gradients is both robust and efficient. 
\end{abstract}

\maketitle

\section{Introduction}
The availability of increasingly sensitive cosmological and astrophysical data is accompanied by further 
scrutiny on General Relativity (GR) as the theory of gravity governing the universe (e.g.~\cite{Will:2014bqa,Freire:2012mg,Baker:2012zs,Yunes_Siemens13,Berti_etal15,Yunes_etal16}). 
GR continues to successfully meet tests so far enabled by such data; indeed, examples include GR's 
successful role in $\Lambda$CDM (e.g.~\cite{Aghanim:2018eyx}), 
binary pulsar timing (e.g.~\cite{Freire:2012mg}), and gravitational waves from compact
binary systems (e.g.~\cite{Abbott_etal16}).
Nevertheless GR should eventually show cracks\footnote{Though a part of the theoretical physics community
argues this is already the case due to the need for dark matter and dark energy.} and the search for possible deviations together with their physical implications, 
are greatly enhanced through theoretical analysis of putative extensions to GR. 
Of particular interest is the highly dynamical, strongly gravitating regime which, arguably, 
presents the likeliest scenario to detect such deviations.

The analysis in such regimes is however hindered by several facts. First, it is yet unclear which specific
theory (theories) one should focus on. Many theories have been proposed motivated by quantum gravity
ideas, explorations of specific violations of fundamental principles of GR, alternatives to
dark matter/dark energy, etc. While many of them are certainly appealing from academic reasons, 
no subset of theories has yet arisen as a preferred one. 

Second, and at a mathematical level, many such theories do not lend themselves to defining well-posed problems 
--especially in the regime of interest. That is, given suitable initial conditions, a unique
solution can be determined which depends continuously on such conditions~\cite{jH02}. Indeed, 
with the exception of just a few theories (within the subset of scalar-tensor or scalar-vector-tensor theories), 
which have been explored nonlinearly in compact binary 
mergers~\cite{Healy:2011ef,Barausse:2012da,Sagunski:2017nzb,Hirschmann:2017psw},
lack of well-posedness presents a severe obstruction to the study of such theories. Of particular relevance
is the understanding of compact objects, their stability and behavior in compact binary mergers, as well
as cosmologies near the big-bang/or bounce, both of which are nonlinear regimes that naturally involve 
relativistic speeds and strong gravitational fields.

Of course, mathematical difficulties have been identified before in some theories and to different degrees. For instance, efforts have been directed towards avoiding so-called Ostrogradski's 
instabilities (e.g.~\cite{Woodard:2006nt,tCmFeLaT13,deRham:2014zqa,Solomon:2017nlh}). Such instability,
when present, invariably leads to ill-posed problems. Unfortunately, even in theories free of
such instability, well-posedness is far from guaranteed, as several aspects still need to be determined (see,
e.g.~\cite{Sarbach:2012pr}).

One traditional way to assess well-posedness --which relies on energy estimates-- involves understanding the 
structure of the (principal part of the)  evolution equations and assess whether they are: (i) symmetric/strongly
hyperbolic, (ii) merely weakly hyperbolic or even if (iii) 
the system can transition from hyperbolic to elliptic within
the domain of interest. Property (i), together with appropriate initial and boundary conditions, 
is essentially sufficient to guarantee well-posedness, irrespective of the lower-order terms of the equations.
Property (ii), however, requires a careful analysis of lower-order terms and, with the exception of rare
circumstances and in rather simple problems, ill-posedness typically follows in this case. As examples, one
could mention that in Horndenski theories, weak hyperbolicity in all but a narrow corner has been recently demonstrated~\cite{Papallo:2017qvl} and similarly concerning aspects have been discussed 
for gravitational Dynamical Chern Simons theory~\cite{Delsate_etal15}. Also, numerical simulations in theories
invoking a Vainshtein-like mechanism have explicitly shown pitalls presented by some of the mathematical 
roadblocks~\cite{Brito:2014ifa} aluded.
Last, property (iii) indicates there exist severe obstructions to determining the solution
beyond where/when the hyperbolic/elliptic transition takes place.

A second hurdle, even with a strongly/symmetric hyperbolic system, is the fact that nonlinearities --in the
principal part-- typically 
imply that characteristics can cross. Once they do, uniqueness of the solution is lost --and hence well-posedness. 
Such difficulties are already encountered at the level of hydrodynamical equations but, in such cases, further requirements
--the so called Rankine–-Hugoniot conditions-- ~\cite{Rankine01011870,Hugoniot} are invoked to single out a unique solution and restore
well-posedness. As far as we know, the analog of such conditions in gravitational theories has yet
to be developed. Finally, there exist theories for which partial differential equations theory is still to
be developed. An example of such a case is given by the ``simple looking'' 
equation $\Box \phi = \lambda (\Box \phi)^p + ... $ with $p>1$, for which there is no mathematical guidance on how to even formally define an initial value problem.

The above discussion paints, at first sight, a rather bleak picture with regards to the degree to which 
most extensions to GR can be thoroughly analyzed. But it might not be the case in practice. 
On one hand, in perturbative regimes
--i.e. cosmology applications-- the study of the linear perturbations with respect to a preferred solution
allows from introducing an ultraviolet cutoff that circumvents many of the aforementioned
problems. Thus, specific predictions can be drawn to test for possible deviations from GR and, with them
hopefully draw lessons for the full theory\footnote{Note however this assumes {\em linearization stability}
of the theory. That is, solutions to the linearized problem are consistent with those of the full
problem linearized. In the case of General Relativity, the conditions for this being
the case were discussed in~\cite{fischer1973}.}. At the nonlinear level however, the situation is more 
uncertain. In such case, nonlinearities could be responsible for runaway energy cascades to the UV, rendering
effective cutoffs delicate to impose without severely affecting the physics.
Moreover, as such regimes require numerical simulations --which at
the level of truncation error typically do source all frequencies allowed by the computational grid--
sensitivity/uncontrolled growth at high frequencies represents a significant practical shortcoming
standing in the way of exploring such regimes. Last, it is important to keep in mind that often extensions
of GR are obtained from an effective field theory approach. A strong direct energy cascade to the UV takes
the solution away from the regime of applicability of the theory and conclusions drawn in such a case need 
not be connected
to the true phenomenology from the putative parent theory from which the EFT is, in principle, derived.

The aforementioned shortcomings are sufficiently delicate that, in a sense, they have prevented significant
advances in the study of extensions to GR in nonlinear regimes. However, the detection of gravitational waves by
LIGO/VIRGO, together with our current rather robust understanding of compact binary mergers in General
Relativity, give hope to shedding light on this enterprise. 
Namely, current observations are consistent with no significant energy cascade to the UV in these systems.
Remarkably, GW150914 has, in a rather spectacular fashion, allowed observing such waves even without invoking
a specific theory!~\cite{Abbott:2016blz}. 
Whatever the true theory of gravity is, it seems not to significantly alter the behavior 
from that of GR --at least in the context of binary mergers-- 
and putative higher-order corrections to GR seem to remain relatively small.

Furthermore, this lack of a strong direct energy cascade to the UV in the context
of compact binary mergers is also what is observed in fully nonlinear simulations within 
General Relativity (see for instance~\cite{Lehner:2014asa,Choptuik:2015mma,Duez:2018jaf} 
and references cited therein).

These observations present both opportunities and
challenges. An exciting opportunity is to study extensions of GR which, as mentioned,
even with pathologies might allow one to devise suitable techniques to effectively control them 
while still capturing the essence of GR departures. After doing so, the challenge is to obtain sufficiently accurate predictions that can be tested or constrained through ever more sensitive detections.

In the current work, we concentrate on assessing proposed techniques to deal with extensions to GR. As the
techniques we study are sufficiently general, we will refrain from making contact with any specific extension.
Instead, we find it more informative to study a toy model which captures essentially all problems
commonly found in extensions to GR. Crucially, being ``UV-complete'' allows for checking to what extent
the solution to the EFT problem stays close to that from the original theory. This is a luxury not yet
afforded by extensions to GR.

To fix ideas and briefly anticipate the strategies we will consider, let us
assume the equations from a given theory can be schematically expressed as
$G(g) = \lambda S(g)$. Here, and for our current purposes, $G$ stands for the Einstein tensor and $S$ some nonlinear,
perhaps higher derivative operator, $\lambda$ is a coupling parameter considered small and $g$ the metric 
tensor\footnote{Also, further coupled fields can be considered, with their own equations of motion}.

One of these strategies, often referred to as {\em ``reduction of order''}, extends 
a successful technique from ordinary differential equations~\cite{FORD1991217} to the less certain partial differential
equation arena. In this technique, one first solves for $G=0$ (i.e. GR); the solution to this problem $g^{(0)}$ is used
to evaluate the ``source'' $S(g^{(0)})$ and a new solution $g^{(1)}$ is obtained from $G(g^{(1)})=\lambda S(g^{(0)})$.
This process is iterated to a certain desired tolerance of the difference between two subsequent solutions,
i.e. $|g^{(i)}-g^{(i-1)}| < \epsilon$.

The second strategy, which we refer to as {\em ``fixing equations''} introduced 
in~\cite{Cayuso:2017iqc},  extends a suggestion by Israel and Stewart to address ill-posedness
of relativistic hydrodynamics. Here, a new variable $\Pi$, and its
evolution equation (of the form\footnote{A second-order form with the
Box operator is also possible, see~\cite{Cayuso:2017iqc}}  $\tau \Pi_{,t} = -\Pi + \lambda S$), 
are introduced such that $\Pi$ asymptotes to $\lambda S$ (in a timescale determined by $\tau$) and 
controls higher gradients
of $\Pi$. Thus it remains consistent with the original system in the IR while restraining a runaway 
direct energy cascade to the UV.

Both the above prescriptions rely on the assumption that the physical system, for the regime
of interest, does not naturally transfer energy to shorter wavelengths in a significant way.
The former as it assumes $S$ stays as a small correction and the latter as it controls 
the shorter wavelengths. As mentioned, observations of gravitational waves by LIGO/VIRGO seem to support such an
assumption in at least some regimes\footnote{As well as fluid-gravity duality and related arguments, see~\cite{Cayuso:2017iqc}} and thus, invigorates examining these strategies in detail.
This is the purpose of this work.

\section{Model}
We will apply our ideas to a toy model for a complex scalar field 
$\phi$ with U(1) symmetry, as introduced in \cite{Burgess:2014lwa}. Consider the following action:
\begin{equation}
S = -\! \int_{\mathbb{R}^4} \mathrm{d}^4x \left[ \left( \partial_\mu \phi^* \right) \left( \partial^\mu \phi \right) + V(\phi^*\phi) \right] \,,
\label{Action}
\end{equation}
with a potential $V(\phi^*\phi)$ given by
\begin{equation}
V(\phi^*\phi) = \frac{\lambda}{2}\left( \phi^*\phi - \frac{v^2}{2} \right)^2 \,.
\end{equation}
At the classical level we will regard the action given by eq.\ \eqref{Action} as defining a ``UV-complete'' theory. 
Following \cite{Burgess:2014lwa}, an EFT from this action parameterized by the mass scale
$M$ can be derived. To this end, one introduces new real fields $\rho$ and $\theta$ and exploits that the minimum of the potential lies
at $\phi^*\phi= {v^2}/{2}$ to write
\begin{equation}
\phi(x) = \frac{v}{\sqrt{2}}\left[ 1 + \rho(x) \right]\exp[i\theta(x)] \,.
\end{equation}
Then the action given by eq.\ \eqref{Action} can be expressed as
\begin{equation}
\begin{split}
\frac{S}{v^2} ={}& -\! \int_{\mathbb{R}^4} \mathrm{d}^4x \, \mathcal{L}(\rho,\theta) \,,
\label{ActionRhoTheta1}
\end{split}
\end{equation}
with a Lagrangian density given by
\begin{equation}
\mathcal{L}(\rho,\theta) \equiv \frac{1}{2} \left( \partial_\mu \rho \right) \left( \partial^\mu \rho \right) + \frac{1}{2} \left( 1 + \rho \right) \left( \partial_\mu \theta \right) \left( \partial^\mu \theta \right) + V(\rho) \,,
\label{Lagrangian}
\end{equation}
and the potential $V$ is defined as
\begin{equation}
V(\rho) = \frac{M^2}{2} \left( \rho^2 + \rho^3 + \frac{\rho^4}{4} \right)\,.
\label{Potential}
\end{equation}
The spectrum of fluctuations  about the vacuum described by eq. \eqref{Lagrangian} includes a 
massive field $\rho$ and a massless Goldstone boson $\theta$. The resulting equations of motion are,
\begin{eqnarray}
\Box \rho &=& \left( 1 + \rho \right) \left( \partial_\mu \theta \right) \left( \partial^\mu \theta \right) + V'(\rho) \, , \label{RhoEL} \\
\Box \theta &=& -2 \, (1+\rho) \, \rho_{,\mu} \partial^{\mu} \theta \label{ThetaEL} \, .
\end{eqnarray}
Now, when $M$ is large compared with the energies of interest, the field $\rho$ can
be integrated out and capture, through an effective action its leading-order
effects on $\theta$. As described in~\cite{Burgess:2014lwa}, this leads to
\begin{equation}
\frac{S}{v^2} \simeq -\! \int_{\mathbb{R}^4} \mathrm{d}^4x \, {\widetilde{\mathcal{L}}(\theta,\partial_\mu \theta,\partial_\mu \partial_\nu \theta)} \, ,
\label{EffectiveAction4}
\end{equation}
where, to ${\cal O}(M^{-4})$,
\begin{eqnarray}
\widetilde{\mathcal{L}} &\equiv & \,\, \frac{1}{2} \left(\partial_\nu \theta \right) \left(\partial^\nu \theta\right) - \frac{1}{2M^2}{\left[ \left( \partial_\nu \theta \right) \left( \partial^\nu \theta \right) \right]^2}
 \nonumber \\
& &+ \frac{2}{M^4} \left( \partial_\mu \partial_\nu \theta \right) \left( \partial^\mu \partial^\sigma \theta \right) \left( \partial^\nu \theta \right) \left( \partial_\sigma \theta \right) \, .
\label{LagrangianTilde}
\end{eqnarray}
The resulting equation of motion for $\theta$ can be expressed as
\begin{eqnarray}
\Box \theta &=& \frac{2}{M^2} \Bigl [ \left( \partial_\nu \theta \right) \left( \partial^\nu \theta \right) \Box \theta + 2 \left( \partial_\mu \partial_\nu \theta \right) \left( \partial^\nu \theta \right) \left( \partial^\mu \theta \right) \Bigr ] \nonumber \\ 
& &+\frac{4}{M^4} \Bigl [ \left( \partial^\nu \partial^\sigma \Box \theta \right) \left( \partial_\nu \theta \right) \left( \partial_\sigma \theta \right) + \left(\partial^\sigma \Box \theta \right) \left( \Box \theta \right) \left( \partial_\sigma \theta \right) \nonumber \\ 
& &+ \left( \partial^\sigma \Box \theta \right) \left( \partial^\nu \theta \right) \left( \partial_\nu \partial_\sigma \theta \right) + \left( \partial^\mu \partial^\sigma \theta \right) \left( \Box \theta \right) \left( \partial_\mu \partial_\sigma \theta \right) \nonumber  \\
& & + 2 \left( \partial^\nu \partial^\mu \partial^\sigma \theta \right) \left( \partial_\mu \partial_\nu \theta \right) \left( \partial_\sigma \theta \right) \Bigr ] \,.
\label{EFT_Equation_General}
\end{eqnarray}
Note the appearance of third- and fourth-order derivatives. In the UV-complete theory, these terms do not arise: we see from eq.\ \eqref{RhoEL} and \eqref{ThetaEL} that the highest derivatives are second-order.

In general, the presence of time derivatives that are higher than second-order in the equations of motion is a strong indication of ill-posedness. This is because such derivatives usually give rise to the so-called Ostrogradski instability: the Hamiltonian is unbounded from below, leading to runaway solutions \cite{Solomon:2017nlh}. Because of the Ostrogradski instability, it has been argued that keeping all time derivatives in the action at second-order and below should be an important restriction for any physical theory.  However, as illustrated in~\cite{Cayuso:2017iqc}, higher-order spatial derivatives are also
problematic.


Importantly, as discussed in~\cite{Solomon:2017nlh}, it is possible to address, to some extent, certain potential sources of ill-posedness arising from EFTs. 
Indeed, through field redefinitions higher-order derivatives can be removed, or pushed to higher orders.
In our case, one can use a field redefinition to push the higher-order time derivatives in eq.\ \eqref{EFT_Equation_General} to $\mathcal{O}\left(M^{(-6)}\right)$. 
As described in \cite{Solomon:2017nlh}, let
\begin{equation}
\theta \rightarrow \theta + \frac{2}{M^4}\left( \partial_\mu \partial_\nu \theta \right) \left( \partial^\mu \theta \right) \left( \partial^\nu \theta \right) \,.
\end{equation}
One can then rewrite the EFT action given by eq.\ \eqref{EffectiveAction4} as
\begin{equation}
\begin{split}
\frac{S}{v^2} &\simeq -\! \int_{\mathbb{R}^4} \mathrm{d}^4x \, \Biggl \{ \frac{1}{2} {\partial_\nu \theta \,
 \partial^\nu \theta } - \frac{1}{2M^2}{\left[ \partial_\nu \theta \, \partial^\nu \theta \right]^2} \\
& - \frac{1}{M^4}{\partial_\sigma \theta \, \partial^\sigma \theta} \left [ \partial_\mu \partial_\nu \theta \, \partial^\mu \partial^\nu \theta - \left[ \Box \theta \right]^2 \right ] \Biggr \} \, .
\end{split}
\label{EffectiveAction6}
\end{equation}
The equation of motion that follows from the action given by eq.\ \eqref{EffectiveAction6} is 
\begin{eqnarray}
\Box \theta &=& \frac{2}{M^2} \Bigl [ \left( \partial_\nu \theta \right) \left( \partial^\nu \theta \right) \Box \theta + 2 \left( \partial_\mu \partial_\nu \theta \right) \left( \partial^\nu \theta \right) \left( \partial^\mu \theta \right) \Bigr ] \nonumber \\
& & + \frac{2}{M^4} \left [ 3\,\Box \theta \left( \partial_\mu \partial_\nu \theta \right) \left( \partial^\mu \partial^\nu \theta \right) - \left( \Box \theta \right)^3 \right. \nonumber \\
& & ~~~~~~~~~ - \Bigl. 2\left( \partial_\mu \partial_\nu \theta \right) \left( \partial_\sigma \partial^\mu \theta \right) \left( \partial^\sigma\partial^\nu \theta \right) \Bigr] \, .
\label{Other_EoM}
\end{eqnarray}
Note the higher-order derivatives cancelled out, thus sidestepping the Ostrogradski instability. A related
approach to remove higher derivatives is to consider adding counterterms in the action~\cite{Chen:2012au}.
At this point, we find it important to stress once again that {\em the absence of this instability 
does not guarantee well-posedness}. Other problems might be in the way still like: lack of uniqueness,
uncertain character of the resulting equations, runaway cascade to the UV, etc. \\

In the next section, we will illustrate techniques proposed to deal with such issues in the nonlinear regime.
Necessarily, this will require a numerical treatment. We
stress, however, that the issues we are dealing with are analytical in nature, but they have strong implications on
the way we can explore through numerical means the targeted regimes.


\section{Assessment Framework}
As discussed, our regime of interest is nonlinear, strongly gravitating and highly dynamical. In this enterprise,
numerically implementing the equations will be required to at least gain some intuition of the dynamics
of the problem under study. As a result, analytical considerations will intertwine with numerical ones. 

We first provide an outline of several techniques that can be used for obtaining the numerical solutions to the EFT equations of motion. For simplicity, we limit ourselves to one spatial dimension and, for concreteness we will
restrict to terms up to $\mathcal{O}(M^{(-2)})$. 
Unless stated otherwise, however, our techniques are equally applicable to the EFT equation of motion with higher-order terms. The truncation of terms of order ${\cal O}(M^{(-4)})$ and above is simply for convenience.

In one spatial dimension, eq.\ \eqref{EFT_Equation_General} to ${\cal O}(M^{(-2)})$ can be written as
\begin{eqnarray}
\left(\theta_{xx} - \theta_{tt} \right)\Bigl [-1 + \frac{2}{M^2} \left( \theta_x^2 - \theta_t^2 \right) \Bigr ] & & \nonumber \\
+ \frac{4}{M^2} \left(\theta_{xx} \theta_x^2 - 2\theta_{xt}\theta_x\theta_t + \theta_{tt}\theta_t^2 \right) &=& 0 \,.
\label{EffectiveEoMSpecial}
\end{eqnarray}
%
%
%
Note that eq.\ \eqref{EffectiveEoMSpecial} contains no more than two derivatives per field. Furthermore,
it is also free of further complications that do appear at higher orders. In particular, second derivatives of
$\theta$ (the principal part of the system) appear multiplied {\em at most} by first derivatives of the field $\theta$.
Consequently, the equation still has a known PDE hyperbolic character; however, standard arguments indicate characteristics
will generically cross~\cite{john2013partial}. This will lead to non-uniqueness and the loss of differentiability of
the solution which, in turn, implies that one abandons the regime of applicability of the EFT expansion. 
Nevertheless, until this happens, eq.\ (\ref{EffectiveEoMSpecial}) can be
directly solved numerically and we will do so to compare with those obtained through approximate strategies.
Similarly,  we can also obtain the solution  of $\{\rho,\theta\}$ in the full theory 
(eq.\ \ref{RhoEL}, \ref{ThetaEL}) and such solutions will be used as the base ones to compare against.
In what follows we outline the basic implementation of each option.

\subsection{``UV-Complete'' Problem}
Since we restrict to the classical regime, the original problem described 
by eq.\ \eqref{RhoEL} and \eqref{ThetaEL}
can be formally studied for any frequency. We thus refer to this as the ``UV-complete'' problem, and solutions obey,
\begin{align}
\label{UVCompleteEq1}
\rho_{tt} - \rho_{xx} &= \left( 1 + \rho \right) \left( \theta_t^2 - \theta_x^2 \right) - \frac{M^2}{2} \left( 2\rho + 3\rho^2 + \rho^3 \right)\,,\\
\label{UVCompleteEq2}
\theta_{tt} - \theta_{xx} &= \frac{2}{1+\rho}\left( \rho_x \theta_x - \rho_t \theta_t \right) \,.
\end{align}
%
For practical reasons, we will write this system in first-order form, by introducing the variables,
$\{f \equiv \theta_x\, , \, g \equiv \theta_t \,,
u \equiv \rho_x\,, \,v \equiv \rho_t \,\}.$
Then eq.\ \eqref{UVCompleteEq1} and \eqref{UVCompleteEq2} can be written as the following six coupled equations:
\begin{align}
f_t &= g_x\,, \\
g_t &= f_x + \left( \frac{2}{1+\rho} \right) \left( f u - g v \right) \,, \\
u_t &= v_x\,, \\
v_t &= u_x + \left( 1 + \rho \right) \left( g^2 - f^2 \right) - \frac{M^2}{2}\left( 2\rho + 3\rho^2 + \rho^3 \right)\,, \\
\rho_t &= v \,, \\
\theta_t &= g \,.
\end{align}
Notice that there are no terms with fields multiplying the principal part (first derivatives) of the equations; the
system is {\em linearly degenerate} and will not lead to discontinuities from smooth initial data.

\subsection{EFT: $\boldsymbol{\mathcal{O}(M^{(-2)})}$ Truncated Theory}
In terms of the variables
$\{ f \equiv \theta_x \, \text{and} \, g \equiv \theta_t \,\}$ eq.\ \eqref{EffectiveEoMSpecial} can be reduced to the following three coupled first-order equations:
\begin{align}
f_t &= g_x \, ,
\label{EffectiveEoMSpecial1} \\
g_t &= f_x \left [ 1 - \frac{4\left( f^2 + g^2 \right)}{M^2 - 2 f^2 + 6 g^2} \right ] + g_x \left( \frac{8 f g}{M^2 - 2 f^2 + 6 g^2} \right) \,, 
\label{EffectiveEoMSpecial2} \\
\theta_t &= g \, .
\label{EffectiveEoMSpecial3}
\end{align}
As already anticipated, eq.\ \eqref{EffectiveEoMSpecial2} contains terms with fields multiplying the principal parts of the equation. The equation is now {\em truly nonlinear} and discontinuities will generically
develop from smooth initial data~\cite{john2013partial}. Further, note we can consider this method 
at this order only, as higher-order ones will contain higher derivatives which will require special treatment. Thus, the possibility of
``direct integration'' will be of limited application. We will primarily contemplate it here for comparison purposes
and to bring attention to some specific issues. In general, other options, like the ones discussed next, will
be required.\\

The next two strategies are envisioned to address these shortcomings, and we will assess their behavior 
with some examples in the next section.

\subsection{Approach 1: ``Fixing the Equations"}
The first strategy to study the EFT equation of motion is based on the methodology introduced 
in~\cite{Cayuso:2017iqc}, which, in turn, is inspired by the Israel-Stewart formulation of relativistic viscous 
hydrodynamics \cite{Israel_Stewart79} (see also, e.g. \cite{Baier_etal08}). The key idea in this approach is to control terms in the equation of motion that effectively contain higher-order gradients, thereby preventing a runaway towards the UV. In the current
problem, this can be achieved in the following way.
Modify the $\mathcal{O}(M^{(-2)})$ (or higher) EFT equation by introducing a new 
variable, $\Pi$ with its own evolution equation as,
\begin{align}
\Box \theta &= \Pi \,, \\
\Pi &=  \frac{2}{M^2} \Bigl [ \left( \partial_\mu \theta \right) \left( \partial^\mu \theta \right) \Pi + 2 \left( \partial_\mu \partial_\nu \theta \right) \left( \partial^\mu \theta \right) \left( \partial^\nu \theta \right) \Bigr ] + \tau \Pi_t \,,
\label{Fixed_General}
\end{align}
where $\tau$ is an external timescale. Note that if $\tau$ is negative, then the effect of the extra term in eq.\ \eqref{Fixed_General} is to drive $\Pi$ to
\begin{equation}
\frac{2}{M^2} \Bigl [ \left( \partial_\mu \theta \right) \left( \partial^\mu \theta \right) \Box \theta + 2 \left( \partial_\mu \partial_\nu \theta \right) \left( \partial^\mu \theta \right) \left( \partial^\nu \theta \right) \Bigr ] \,,
\end{equation}
thus recovering the original equation. The above eqn.\ \eqref{Fixed_General} is one equation that achieves 
the desired goal,
however, as noted in~\cite{doi:10.1063/1.530958}, as long as higher gradient terms remain small, different equations
can be envisaged, and the choice is largely irrelevant if no significant direct cascade of energy takes place.
In one dimension, eq.\ \eqref{Fixed_General} can be expressed as
\begin{eqnarray}
\theta_{xx} - \theta_{tt} &=& \Pi \,, \\
\Pi &=&  \tau \Pi_t + \frac{2}{M^2} \Bigl [ \left( \theta_x^2 - \theta_t^2 \right) \Pi \nonumber \\ & & 
+ 2 \left( \theta_{xx} \theta_x^2 + \theta_{tt} \theta_t^2 - 2\theta_{xt} \theta_x \theta_t \right) \Bigl ] \,.
\end{eqnarray}
Again, introducing
$\{ f \equiv \theta_x \, \text{and} \, g \equiv \theta_t \,\}$,
one ends up with the following system of coupled equations:
\begin{eqnarray}
f_t &= &g_x \,, 
\label{Fixed_Special_1} \\
g_t &=& f_x - \Pi \,, 
\label{Fixed_Special_2} \\
\theta_t &=& g \,, 
\label{Fixed_Special_3} \\
\Pi_t &=& \frac{1}{\tau} \Biggl \{ \Pi - \frac{2}{M^2} \Bigl [ \left( f^2 - g^2 \right) \Pi \Bigr. \nonumber \\
& & + \Biggl. 2 \left( f_x f^2 + \left( f_x - \Pi \right) g^2 - 2 g_x f g \right) \Bigr ] \Biggr \} \,.
\label{Fixed_Special_4}
\end{eqnarray}

\subsection{Approach 2: ``Reduction of Order''}
Once again, we would like to solve the one-dimensional $\mathcal{O}(M^{(-2)})$ EFT equation of motion, as given by eq.\ \eqref{EffectiveEoMSpecial}. Inspired by an approach introduced to deal with the Abraham-Lorentz-Dirac equation
and avoid runaway solutions~\cite{FORD1991217}, an {\em iterative or reduction of order} method is operationally introduced in the following way. One rewrites the equation of motion as
\begin{equation}
\begin{split}
\theta_{xx} - \theta_{tt}&= S(\theta) \,,
\label{EFT_Equation_Source}
\end{split}
\end{equation}
where we define,
\begin{eqnarray}
S(\theta) &\equiv& \frac{2}{M^2} \Bigl [ \left( \theta_x^2 - \theta_t^2 \right)\left(\theta_{xx} - \theta_{tt} \right) 
\nonumber \\
& & + 2 \left(\theta_{xx} \theta_x^2 - 2\theta_{xt}\theta_x\theta_t + \theta_{tt}\theta_t^2 \right) \Bigr ] \,.
\label{Source}
\end{eqnarray}
Then, one regards this equation as one where the leading part is affected by a subleading source $S$. Since
the source depends on the field itself, an iterative procedure is implemented whereby the equation is treated
as $\Box(\theta^{i}) = S(\theta^{i-1})$. The assumption here is that the source $S$ will remain subleading,
and the strategy of decoupling it from the principal part implies, at first sight, that the ``offending'' terms
in $S$ play only a passive role in determining the solution, thus hopefully controlling a runaway behavior
towards the UV (e.g.~\cite{Endlich:2017tqa}). 
Notice that this approach does not introduce further variables, though it does require iterating to
some desired tolerance (the difference between successive solutions).
Notice that the source $S$ has, in particular, second time derivatives of the field $\theta$ and they need
to be evaluated somehow in the iterative scheme. This is no problem at intermediate times, but it is
an issue at (and near) end points of the time interval of interest. We will comment on this in the next section.

\section{Numerical Implementation}

Having discussed the four options we can pursue, i.e.: (i) Full ``UV-complete'' problem,
(ii) Direct integration of the equations truncated to ${\cal O}(M^{-2})$ order (TR), as well as adopting
the (iii) ``Fixed equations'' (F) or (iv) the iterative or ``Reduction of Order'' (RO) approach,
we are now in a position to study and contrast the nonlinear behavior described by each option. 
To this end, we resort to numerical simulations. As a simple setup, we consider a periodic problem
with $x\in[0,L]$ discretized uniformly with spacing $\delta x$. Spatial derivatives are approximated through
Finite Differences through a standard centered fourth-order accurate expression. Time integration is obtained
through the Method of Lines via a Runge-Kutta third-order accurate approximation with $\delta t = \delta x/4$ to
satisfy the CFL condition (for further details see, e.g.~\cite{Calabrese:2003yd,Calabrese:2003vx}).
The initial data is given by,

\begin{equation}
\theta(x, t_0) =
\begin{cases}
10^{-5} (x - 3)^4 (x - 7)^4 & \text{if $x \in [3,7]$} \\
0 & \text{otherwise}
\end{cases}
\label{IC_1}
\end{equation}
and
\begin{equation}
\theta_t(x, t_0) = 0 \,,
\end{equation}
where $t_0 = 0$; for concreteness we adopt $L=10$. 

Notice that for options (iii) and (iv), further information must be specified. From the point of view
the initial value problem, $\Pi$ (as it is a new variable) or $\theta_{,tt}$ (as it is part of the
source) is required at $t=0$.  Providing such information is in and of itself a delicate issue, as it can
aid in removing spurious modes, as discussed in~\cite{Burgess:2014lwa}. In our tests, we adopt 
$\Pi(t=0)=0$, $\theta_{,tt}(t)=0$. In addition, a few ``external parameters'' are required:  \\

\begin{itemize}
\item In case (iii), the parameter
$\tau$. For the most part we will keep it to $\tau=-40/\tau=-4$ (though we will also compare with other values).

\item For (iv), two extra parameters are required. One is the tolerance $\epsilon$ (here chosen to be $=10^{-12}$). 
The other
is the period of {\em physical} time $\Delta T$ over which the integration will be performed. For instance, if
the total physical time of interest is $t_0=0$ to $T$, one could choose to integrate the homogenous problem
$\Box \theta^{(0)} =0$ over this full range (and iterate over it), or over a smaller interval $\Delta T$ and advance
over this interval to cover the full interval of interest.
The introduction of such an interval is natural, and is analogous to what is done, 
for instance in the integration of  the Abraham-Lorentz-Dirac equation \cite{FORD1991217} (see also
\cite{doi:10.1119/1.4897951,Lanir:2017hwl}) through a reduction of order approach. 
There it is found,
as naturally expected, that more accurate solutions are obtained with smaller intervals.

Last, and with respect to this approach, we note a technical but delicate point. Namely, the evaluation 
of second-order (or higher) time 
derivatives that would appear in the application of this scheme when choosing a given $\Delta T$ has an impact on the 
practical interval of time where the solution can be found. Indeed,
a given interval (say $\Delta T$) will not have the final solution (e.g. $\theta^{(n)}$) up to
$t=\Delta T$, as the source could only be accurately evaluated to time $t=\Delta T - \kappa n \delta t$
(with $\kappa$ an integer value related to the accuracy desired in such evaluation, e.g. for fourth order
accuracy $\kappa=2$). As a result, the usable span of time within a given interval is necessarily smaller 
than $\Delta T$.  Besides this technical detail, 
the method can be implemented rather straightforwardly.

\end{itemize}

\section{Results}
In what follows, we will compare errors between solutions obtained with the three approximate
systems of equations \{(ii)-(iv)\} --which we will collectively refer to as EFT solutions-- 
with respect to those obtained with the ``UV-complete'' problem (i). 
Further, we will
take different values of $M$ and study the errors' dependence on this ``coupling'' scale. In the context
of extensions to General Relativity, beyond the Planck length there is, in principle, no 
specific value of coupling naturally expected. Consequently, increasingly sensitive detections would 
help to place bounds on such coupling, provided a clear understanding is obtained for multiple values of it.
It will thus be important to understand the extent to which different methods perform under a range
of couplings within the regime of validity of the EFT approximation. With respect to the initial data
adopted, a reasonable estimate of the size of corrections introduced by the
terms at ${\cal O}(M^{-2})$ is $\approx 2 \,\times 10^5 /M^2$ (estimated
by comparing the magnitude of corrections relative to terms in the principal part of the uncorrected
equations).  In our tests we thus
concentrate on $M=\{0.01, 0.1, 1\}$ to explore a range. All these values would
imply that corrections are, at least initially, small and thus within the regime of validity of the EFT approximation. Naturally, $M=0.01$ will represent the most  demanding case.

To begin, we confirmed the expected convergence and stability of the homogeneous problem ($\Box \theta = 0$) and convergence of solutions obtained with the truncated problem for $M=1$ until $T=5 t_c$. With this implementation,
we next thoroughly explore the behavior of 
solutions obtained with the three EFT
options and contrast their behavior with the ones obtained with the UV-complete problem. 
In particular, we monitor (the $L_2$ norm of the difference) between the solutions and 
compute the relative difference defined as,
\begin{equation}
\textsc{E} \equiv \frac{\left | \left |  \theta_{\text{EFT}} - \theta_{\text{UV}} \right | \right |_{\text{L$_2$}}}{\left | \left | \theta_{\text{UV}} \right | \right |_{\text{L$_2$}}}\,,
\end{equation}
where $\theta_{\text{EFT}}$ denotes each of the EFT solutions and $\theta_{\text{UV}}$ the UV-complete one. \\
\\
For concreteness, we adopt in our production runs $N_x=200$, which shows excellent agreement with the much
finer refined grid of $N_x=1600$ in representative cases.

\begin{figure}[b!]
\centering
\includegraphics[width=2.7in,height=2.2in]{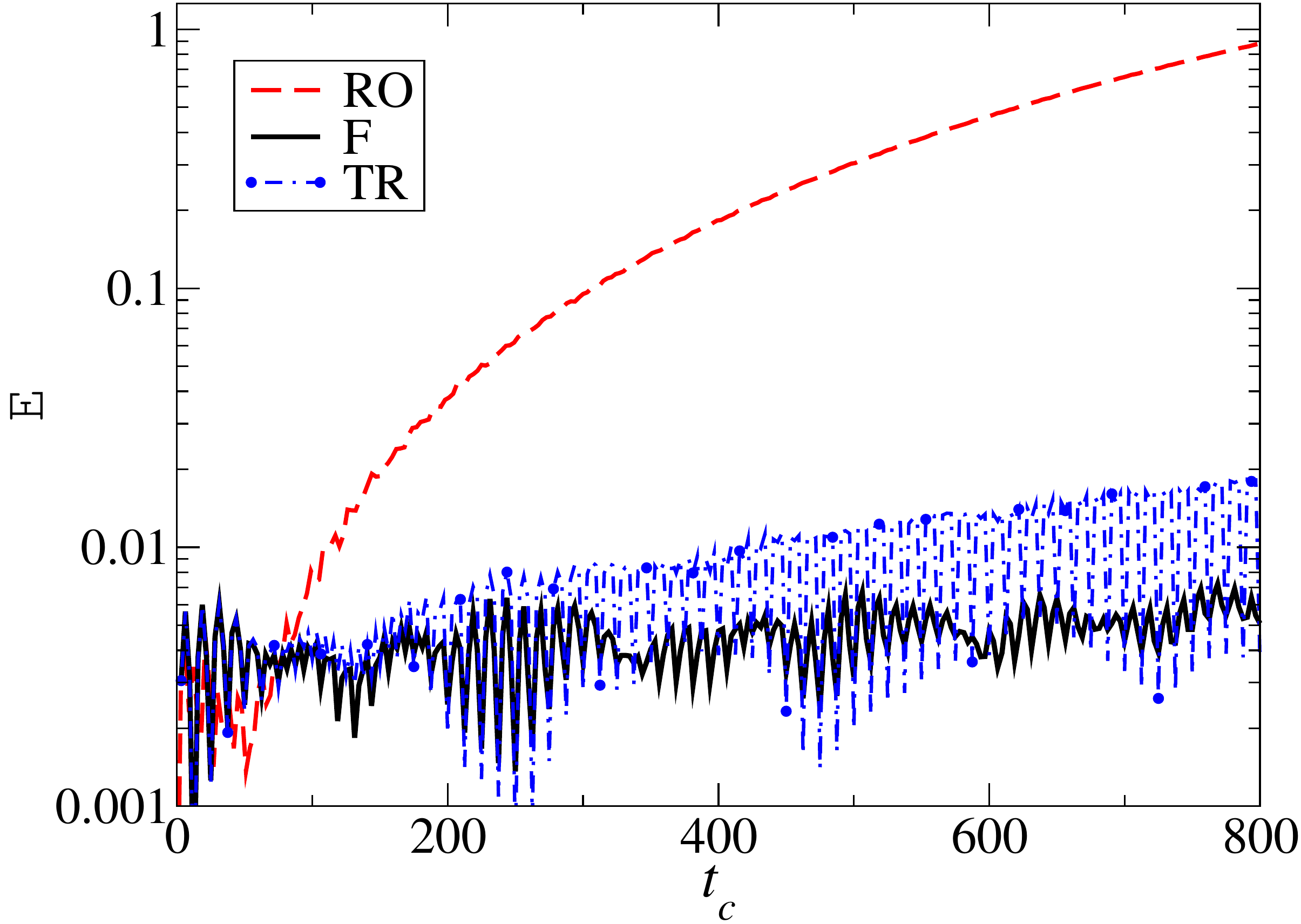}
\includegraphics[width=2.5in,height=2.2in]{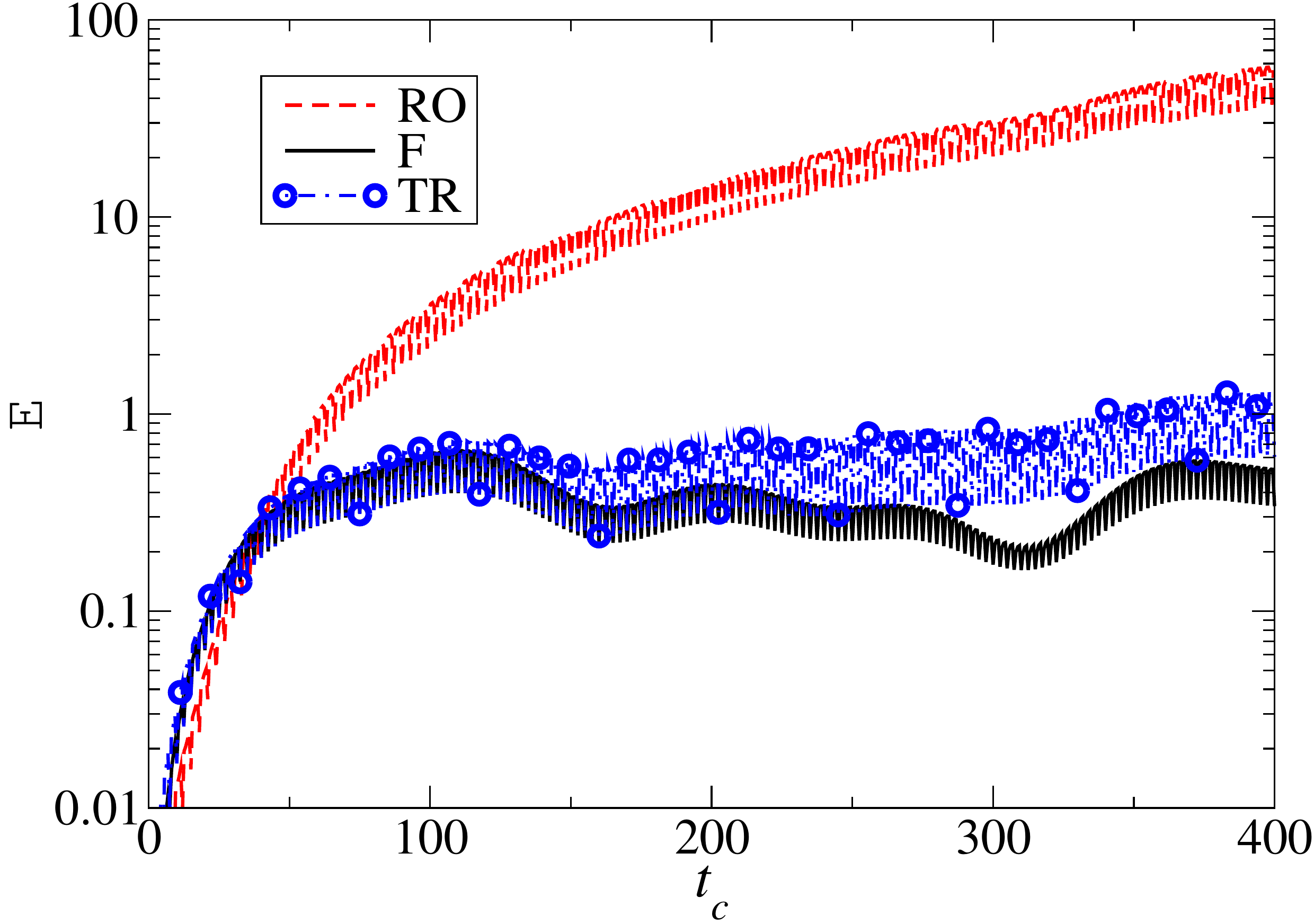}
\caption{Relative difference between the UV-complete and EFT solutions for the three different approaches: 
the truncated (TR), iterative (RO) and fixed (F) schemes, using $M=1$ (top) and $M=0.1$ (bottom).}
\label{Comparison_1}
\end{figure}

Figure \ref{Comparison_1} illustrates $\mathcal{E}(t)$ for the truncated (TR), iterative (RO) and fixed (F) schemes in the  case $M=1$ (left) and  $M=0.1$ (right) versus crossing time ($t_c \equiv t/L$) respectively. For this test, we employ a grid which covers the computational domain with $N=200$ points (and for representative cases 
we compare with solutions obtained with $N=100$ and $N=1600$ to confirm the observed behavior).
For the iterative case, we adopted $\Delta T = 0.015 t_c$, and for the ``fixed'' case $\tau = -4$.
For early times, all methods perform rather well, with the iterative case showing the smallest differences
until $T \simeq 100 t_c$, at which point its behavior gradually worsens when 
compared with the fixed and truncated schemes. Over long integration periods, the fixed method shows
the best behavior, with relative errors of order $\simeq 0.6\%, 50\%$ for $M=1,0.1$ respectively at $T=\{800,400\}$
respectively. The truncated method performs somewhere in between the other two, giving 
errors of order $\simeq 2\%, 100\%$ for $M=1,0.1$ respectively at $T=\{800,400\}$

\begin{figure}[b!]
\centering
\includegraphics[width=2.5in,height=2.2in]{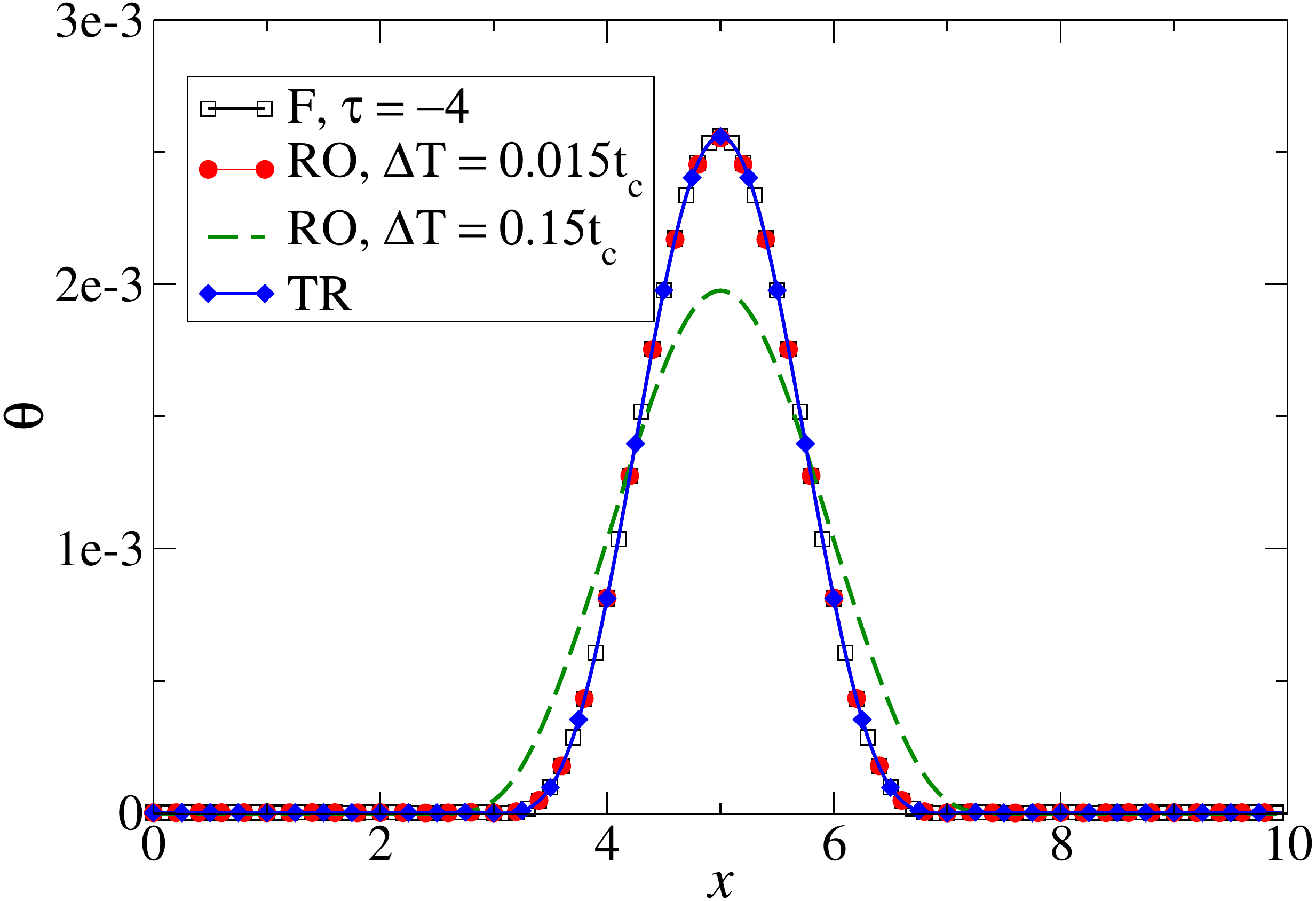}
\caption{Solutions obtained for $M=1$ at $t=19 t_c$. All solutions agree rather well
(for sufficiently small values of $\Delta T$ in the case of the iterative method).}\label{Comparison_solsM1}
\end{figure}

As an illustration of the different solutions at relatively early times, 
figure~\ref{Comparison_solsM1} displays them at $t=19 t_c$. At this stage, and for this coupling, 
all methods produce solutions quite similar to the UV-complete one, with the exception of the iterative 
one if $\Delta T$ is not chosen sufficiently small --recall that the iterative method depends on the value chosen for $\Delta T$.

\begin{figure}[b!]
\centering
\includegraphics[width=2.7in,height=2.2in]{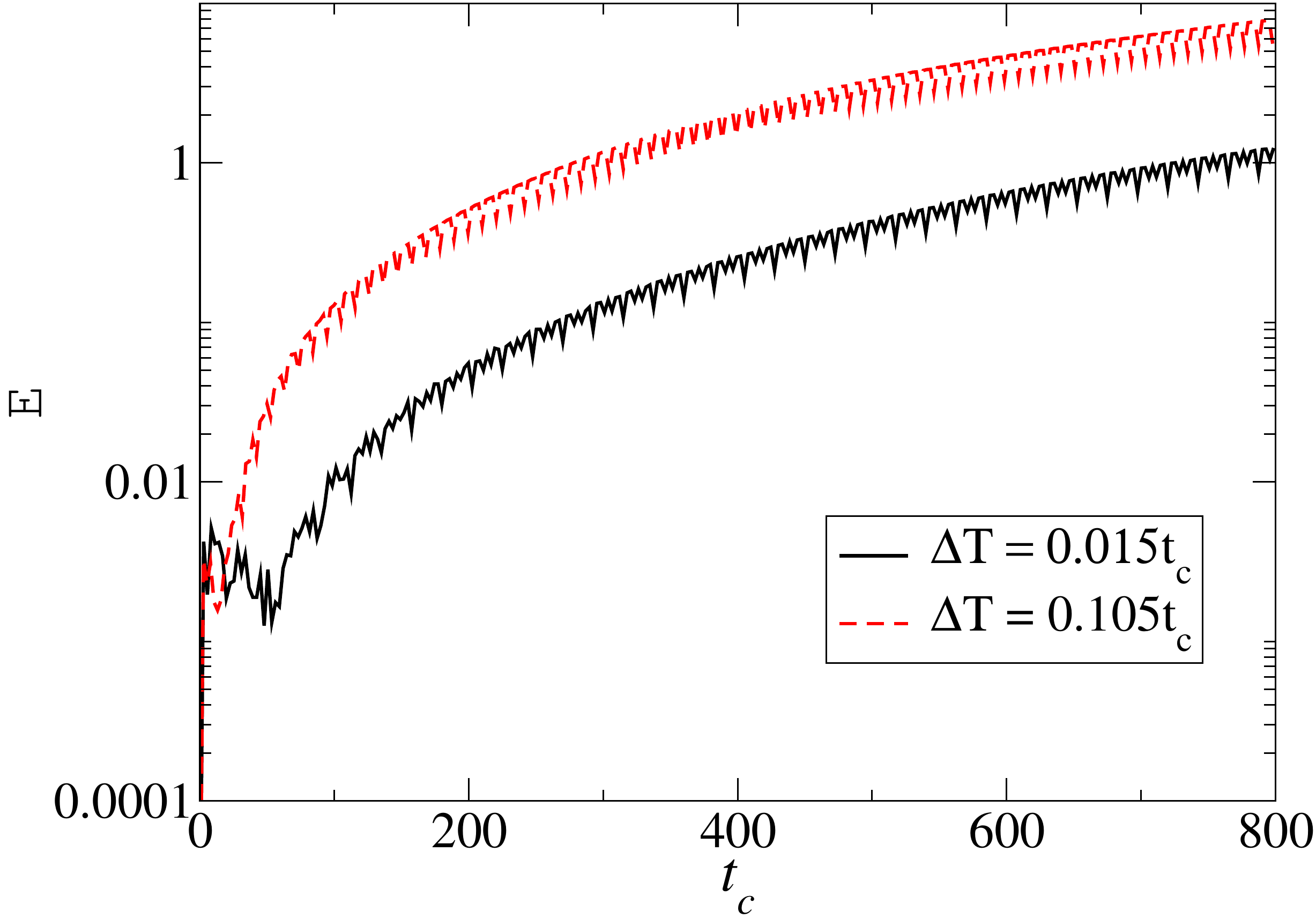}
\includegraphics[width=2.5in,height=2.2in]{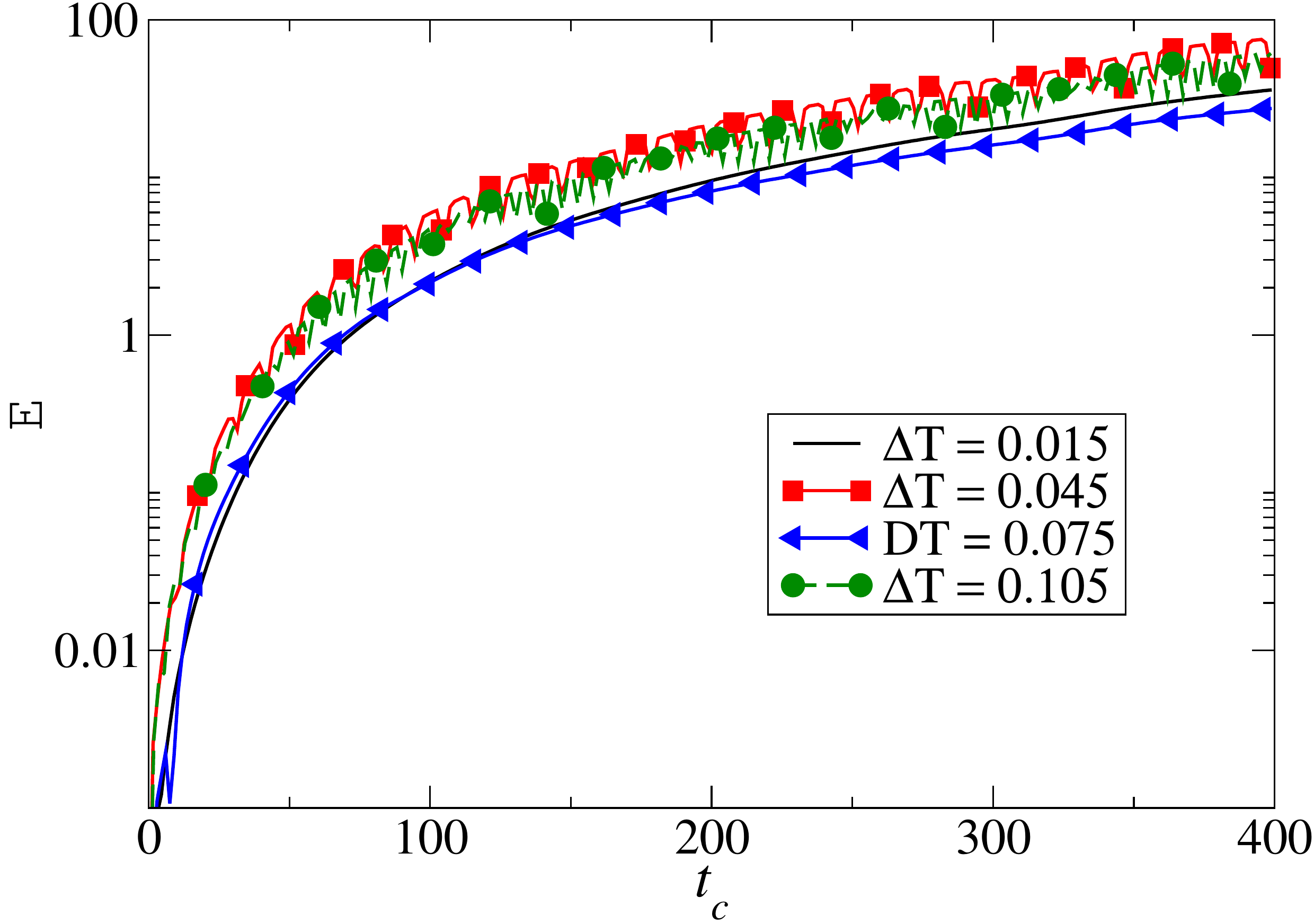}
\caption{Relative difference between the iterative solutions for $M=1$ (top) and $M=0.1$ (bottom) adopting
different values of the iterating physical interval $\Delta T$. For $M=1$ the smallest value of $\Delta T$
performs significantly better --though still showing an exponential growth past $t\simeq 200$--. For $M=0.1$,
while this is still the case, curiously, there seems to be no correlation among the behavior of the solutions
for $\Delta T = \{0.45, 0.75, 1.05\}t_c$. Indeed, the case with $\Delta T = 0.45$ fails to yield
a solution beyond $t\simeq150 t_c$.}\label{Comparison_2}
\end{figure}

To examine more closely the dependency 
of the solutions obtained with the iterative method when varying $\Delta T$, figure \ref{Comparison_2}
presents the relative errors for different values of $\Delta T$ with respect
to the UV-complete case. For the case $M=1$, significantly smaller errors are obtained for $\Delta T = 0.015 t_c$
(i.e. the smallest value adopted) than with $\Delta T = 0.15 t_c$. Additionally, as naturally expected
we find that with  $\Delta T = 0.015 t_c$ satisfying the tolerance requirement involves a couple of iterations
while for $\Delta T = 0.15 t_c$ up to $16$ iterations are involved.
The smallest integration interval also yields the smallest error in the case $M=0.1$, 
but no clear trend is found for larger values of $\Delta T$. We further note that for these larger values
of $\Delta T$ there are a number of instances were the tolerance criteria could not be met, though the
difference achieved for the largest possible number of iterations was still of order $10^{-11}$.

\begin{figure}[b!]
\centering
\includegraphics[width=2.5in,height=2.2in]{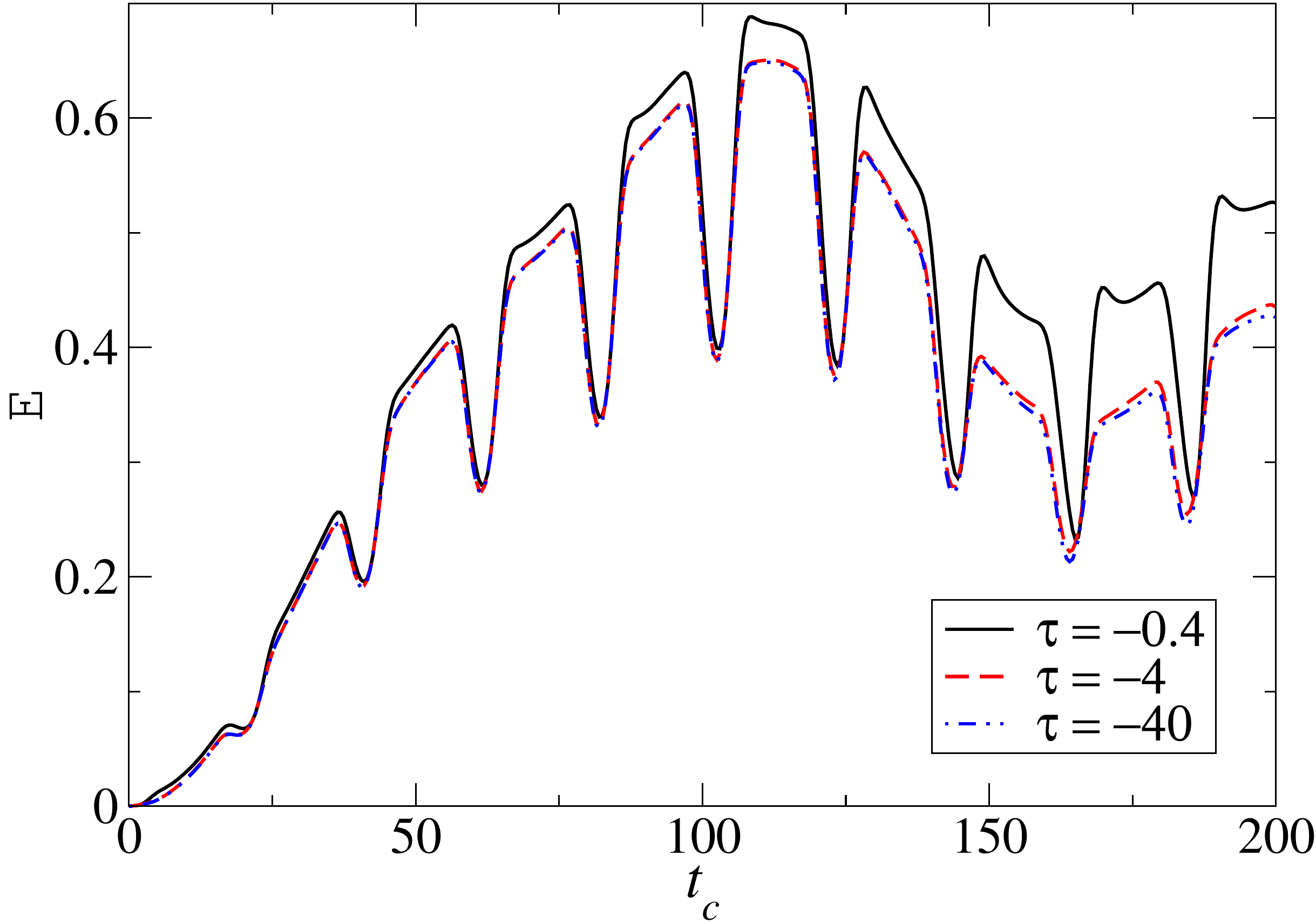}
\caption{Relative norm of the solutions vs time for the case $M=0.1$ for the fixed method
for different values of $\tau$. As $\tau$ is increased, the solutions obtained with the 
fixed method show a better agreement.}\label{Comparison_4}
\end{figure}

The fixed scheme also depends on the external parameter --$\tau$-- and we examine
its sensitivity to varying such parameter. The results are presented in figure~\ref{Comparison_4},
which shows the relative errors for the case $M=0.1$. Upon varying this parameter by two orders of
magnitude, the solutions obtained are rather similar, though for the larger values of
$\tau$ the obtained solutions are closest.

\begin{figure}[b!]
\centering
\includegraphics[width=2.5in,height=2.2in]{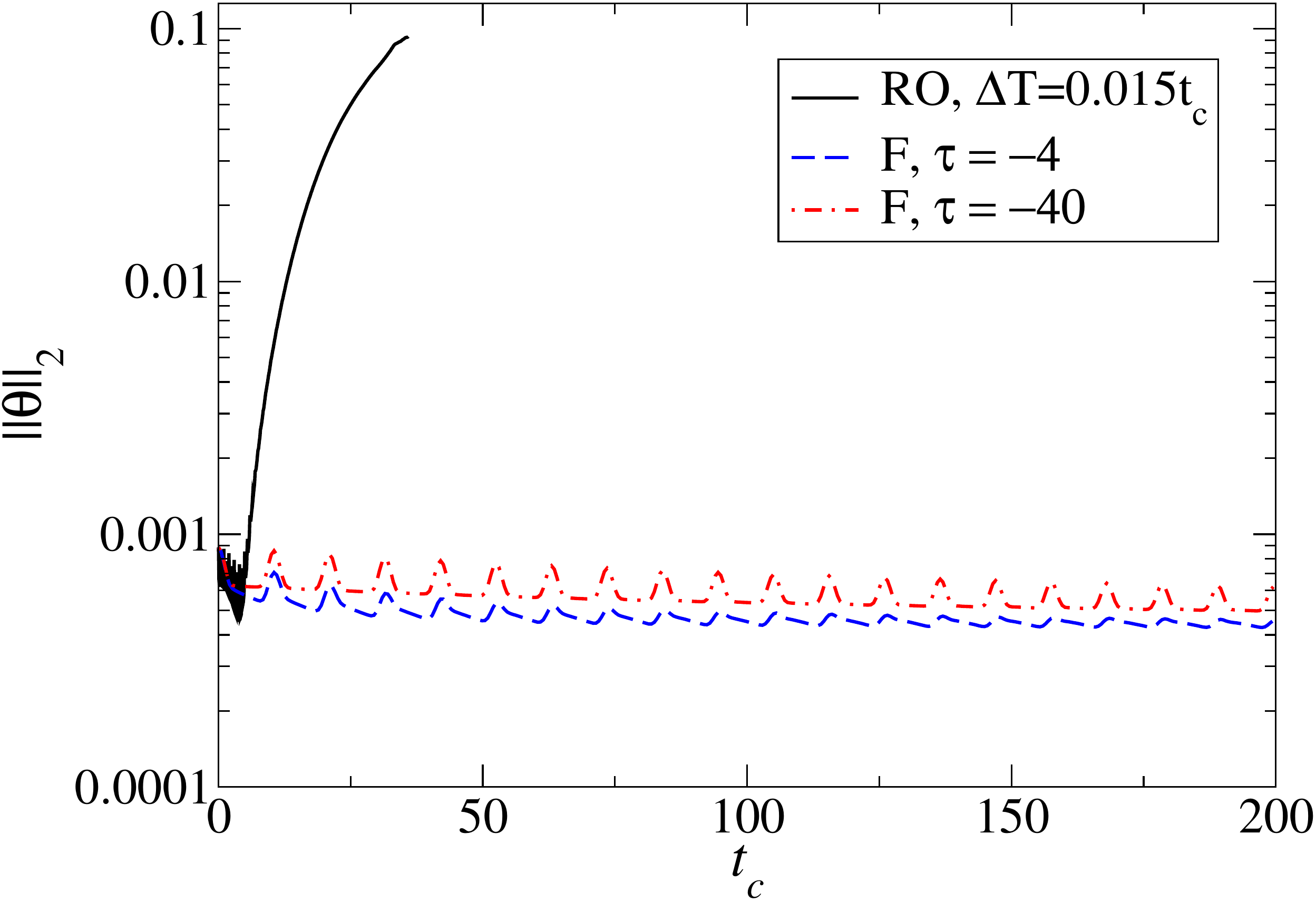}
\caption{Norm of the solutions vs time for the case $M=0.01$. The iterative method fails to
obtain a solution past $t\simeq 35.9 t_c$, exhibiting an exponential blowup. The fixed method
copes with such a small value of $M$ --i.e. a strong coupling value. The solution exhibits a decay
in time however, which is reduced for larger values of $\tau$.}\label{Comparison_3}
\end{figure}

To significantly stress the methods, we also explore solutions for the value 
$M=0.01$, focusing on the iterative and fixed methods in particular. 
Figure~\ref{Comparison_3} shows the norm of the solutions obtained with different
methods for this case. The iterative method copes with such a strong coupling for some time,
but it eventually leads to an exponentially diverging behavior and by $t \approx 35.9 t_c$ can no
longer yield a solution. The fixed method however, stays 
bounded, displaying instead a slow exponential decay that is reduced for larger values of $\tau$.

\begin{figure}[b!]
\centering
\includegraphics[width=2.5in,height=2.2in]{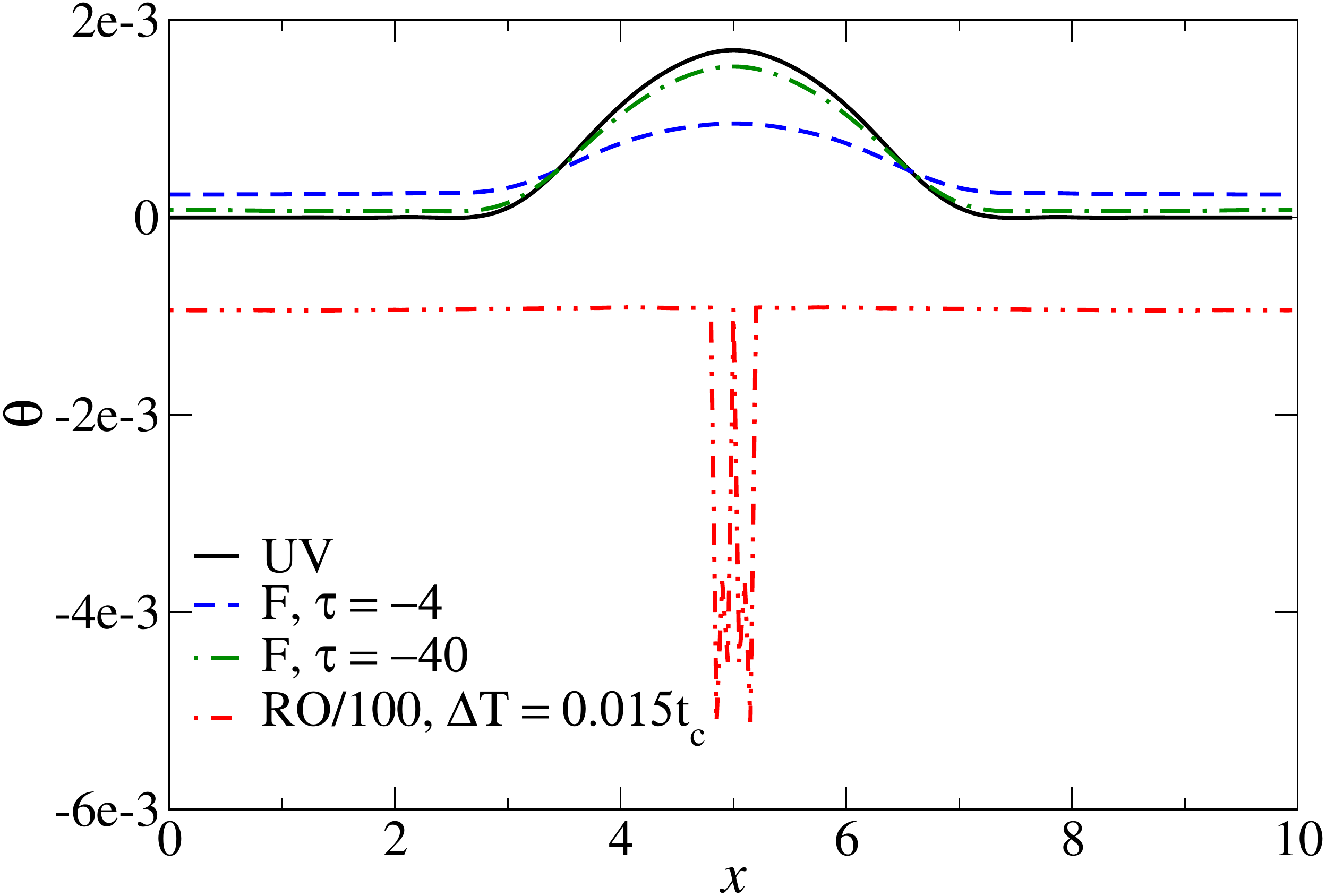}
\caption{Solutions obtained for $M=0.01$ at $t=35.9 t_c$. For this rather strenuous case,
the truncated method fails to provide a solution after a rather short period of time.
The iterative solution (shown here divided by 100 to fit in the plot), even with a small value of $\Delta T$, 
yields a divergent solution after $\simeq 35.9 t_c$.
The fixed solution stays well behaved and for sufficiently large values of $\tau$ reasonably close to the solution
of the UV-complete problem.}\label{Comparison_solsM01}
\end{figure}

We illustrate different solutions for this case in figure~\ref{Comparison_solsM01} at
$t=35.95 t_c$. The solutions obtained with the iterative method clearly diverge at this
stage. The fixed method yields well-behaved solutions, though the larger the value of 
$\tau$, the closer its solution is to that of the UV-complete problem.

\section{Observations and Final Comments}
The unprecedented opportunity to explore gravity in strongly gravitating/highly dynamical regimes has
given further input to explore ideas on extensions of General Relativity and hopes to put them to the test.
However, most such extensions typically involve pathological features that have hindered a thorough 
understanding of their implications, especially in regimes where deviations from GR would be largest.
Recently, several different ideas have been suggested as possible paths forward in this 
endeavour \cite{Endlich:2017tqa,Cayuso:2017iqc,Okounkova:2017yby} and first steps are being 
taken~\cite{Okounkova:2017yby,witekgualtiereprivate,ijjasprivate,pretoriusprivate}. 
In this work, we exploited
a toy-model which --from an EFT point of view-- evidences many of such pathologies, but within a known UV-complete
theory. With this model we assessed possible strategies to study such extensions and illustrated their
strengths and shortcomings. In particular, we find that the approach that ``fixes'' the 
equations~\cite{Cayuso:2017iqc} to
control higher-order gradients performs quite well, both in terms of approximating the UV-complete problem,
the robustness it displays with strong couplings and the computational cost associated to implementing it.
Faithful solutions are obtained for sufficiently large values of $\tau$; though we note that
for very large values, the associated equation for $\Pi$ will become stiff. However, the usage of
IMEX methods~\cite{2010arXiv1009.2757P} and the fact that the stiff term is quite simple, ensures that one can easily implement the equations
without affecting the computational cost.
The reduction of order method appears to perform well for sufficiently weak couplings and small intervals
$\Delta T$. It is relatively straightforward to implement, though it shows a lack of robustness with stronger 
couplings. However, this behavior
could be ameliorated by implementing corrections in a perturbative expansion as introduced recently
in~\cite{Okounkova:2017yby} --if stronger couplings are of interest. We note that due to the iteration required,
its associated computational cost is higher.

Importantly, the faithfulness of all the methods discussed --and in fact the use of EFT to derive a theory
and apply it to a given case-- requires the physical system to not present a strong cascade of energy to the 
UV (or to restrict the solution to the regime where such behavior is not present). 
Otherwise, terms considered and treated as 
small (in the iterative scheme) or controlled (in the fixed scheme) would yield solutions not necessarily
tracking the underlying gravitational theory one wishes to study. Taking GR as an example, many non-trivial
scenarios do not display a strong UV cascade. Examples include scenarios ranging from the nonlinear 
stability of Minkowski~\cite{christodoulou1993global},  and the existence of stable 
stars (e.g.~\cite{1983bhwd.book.....S}) to subtle islands of stability in GR
in the presence of a negative cosmological constant~\cite{Buchel:2012uh,Balasubramanian:2014cja}. 
Of particular interest, the merger of astrophysical compact objects also seems to be one where such a behavior
is satisfied. Both observational
evidence, e.g.~\cite{Abbott:2016nmj,Abbott:2016blz,Abbott:2017} and analytical studies (through simulations),
(see e.g.~\cite{Lehner:2014asa,Duez:2018jaf} and references cited), illustrate a rather simple
behavior where energy is mainly contained in relatively long-wave modes (commensurate with the size of the
objects involved). 

However, we know a strong cascade to the UV does indeed take place in the right regimes.  
For instance,  singularity theorems~\cite{Hawking:1973uf}  do imply a runaway behavior to the UV 
--a runaway that is halted if a black hole forms which introduces
a natural cutoff. It is not known whether this behavior is also present in extensions to GR, we hope this work
helps guide efforts to answer this and related questions through the analysis of solutions to particularly
relevant scenarios. In the current work, the adopted toy-model afforded the luxury of having the ``UV-complete''
solution to compare against. In the gravitational case this is obviously not the case and assessing the faithfulness
of obtained solutions is more delicate. This can be done by evaluating the original (EFT) system of equations and studying its residual. Small values would indicate the solution obtained is within the regime of applicability
of the EFT expansion. Additionally,  one can explore the impact
of the ``extra parameters'' (e.g. $\Delta T, \tau$) and gauge whether the solutions depend on them
in a sensitive manner. Regimes not displaying a significant energy cascade to the UV, or manifesting a natural
cutoff within the regime of applicability of the EFT should pass these tests.

{\it Acknowledgements-} 
We thank C. deRham, M. Johnson, N. Ortiz, O. Reula, M. Okounkova, F. Pretorius, 
L. Senatore, T. Sotiriou, L. Stein, A. Tolley, M. Trodden and R. Wald for helpful discussions.
This research was supported in
part by NSERC, CIFAR and the Perimeter Institute for Theoretical
Physics.  Research at Perimeter Institute is supported by the
Government of Canada through the Department of Innovation, Science and
Economic Development Canada, and by the Province of Ontario through
the Ministry of Research and Innovation.
LL thanks the Mainz Institute for Theoretical Physics (MITP) and
the Centro de Ciencias de Benasque Pedro Pascual for its
hospitality and support during the completion of this work.

\bibliographystyle{utphys}

\providecommand{\href}[2]{#2}\begingroup\raggedright\endgroup

\end{document}